\documentclass[aps,pr,10pt,notitlepage,nofootinbib,superscriptaddress,showkeys,showpacs]{revtex4-2}
\pdfoutput=1
\usepackage{amstext,amsmath,amssymb,amsfonts,bbm,euscript,color,dsfont}
\usepackage[latin1]{inputenc}
\usepackage{epsfig}
\usepackage{hyperref}
\usepackage{amsthm}
\usepackage{subfigure}
\usepackage{color}
\usepackage{multirow}
\usepackage{tikz}

\usetikzlibrary{arrows,backgrounds}

\usepackage{verbatim}

\usepackage{graphicx}

\usepackage{latexsym}

\newcommand{\bea}{\begin{eqnarray}}	
\newcommand{\eea}{\end{eqnarray}}
\newcommand{\be}{\begin{equation}}	
\newcommand{\ee}{\end{equation}}
\newcommand{\beq}{\begin{equation}}	
\newcommand{\eeq}{\end{equation}}

\newcommand{\Z}{{\mathbb Z}}
\newcommand{\C}{{\mathbb C}}

\def\R{\relax\ifmmode {\mathbb R}  \else${\mathbb R}$\fi}
\def\C{\relax\ifmmode {\mathbb C}  \else${\mathbb C}$\fi}
\def\Z{\relax\ifmmode {\mathbb Z}  \else${\mathbb Z}$\fi}
\def\N{\relax\ifmmode {\mathbb N}  \else${\mathbb N}$\fi}
\def\I{\relax\ifmmode {\mathbb I}  \else${\mathbb I}$\fi}

\hypersetup{
colorlinks=true,
citecolor=blue,
citebordercolor=red,
linktoc=all,
linkcolor=blue,
urlcolor=blue
}

\newbox{\ORCIDicon}
\sbox{\ORCIDicon}{\large
	\includegraphics[width=0.8em]{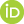}}

\begin{document}

\title{Finiteness of the Yang-Mills-Chern-Simons action in linear covariant gauges \\ 
by taking into account gauge copies}  

\author{Daniel O. R. Azevedo}\email{daniel.azevedo@ufv.br}
\affiliation{Departamento de F\'isica, Universidade Federal de Vi\c cosa,
	Campus Universit\'ario, Avenida Peter Henry Rolfs s/n, 36570-900, Vi\c cosa, MG, Brazil}
\author{Antonio D. Pereira\,\href{https://orcid.org/0000-0002-6952-2961}{\usebox{\ORCIDicon}}} \email{adpjunior@id.uff.br}
\affiliation{Instituto de F\'isica, Universidade Federal Fluminense, Campus da Praia Vermelha, Av. Litor\^anea s/n, 24210-346, Niter\'oi, RJ, Brazil}

\begin{abstract}

In recent years, the effects of removing infinitesimal Gribov copies from the path integral of gauge-fixed Yang-Mills-Chern-Simons theories formulated in three-dimensional Euclidean space have been investigated. Part of the interest resides on the fact that such an elimination of gauge copies introduces a mass parameter, the Gribov parameter, which is relevant when the assumptions of the Faddeev-Popov procedure are not well-grounded. Such a parameter enters the propagator of the gauge field which is topologically massive due to the Chern-Simons term. The resulting action, which eliminates infinitesimal Gribov copies in this context, has been constructed in linear covariant gauges and the interplay between the aforementioned mass-parameters allows for a rich phase diagram in which confining and deconfining signatures are observed in the gauge-field propagator. In the present work, we establish the renormalization properties of such a theory at all orders in perturbation theory by means of the algebraic renormalization framework and show that the removal of infinitesimal Gribov copies does not affect the standard non-renormalization properties of standard gauge-fixed Yang-Mills-Chern-Simons theories, i.e., the theory is finite.
 
\end{abstract}

\maketitle

\section{Introduction \label{Sect:Intro}}

After seventy years of its proposal \cite{Yang:1954ek}, many properties (and their driving mechanisms) of Yang-Mills (YM) theories still lack a first-principles description. Notably, (color) confinement is widely accepted, but a fully dynamical derivation of its origins is still an open problem and different approaches to tackle such a problem have been developed over the past decades, see, e.g., \cite{Greensite:2011zz,Brambilla:2014jmp}. In four dimensions, YM theories are the building blocks of the Standard Model of Particle Physics (SM) and enjoy remarkable properties such as renormalizability and asymptotic freedom \cite{tHooft:1972tcz,Politzer:1973fx,Gross:1973id}. This makes the use of standard perturbative techniques very efficient for sufficiently high energies, where the theory is weakly coupled. Yet the gauge coupling constant grows towards the infrared (IR) and non-perturbative methods become necessary in order to properly access the dynamics of the theory. In fact, perturbation theory breaks down in the IR thanks to a Landau pole which, in this case, is caused by the abuse of the use of perturbation theory in a non-perturbative regime. As it is widely known, the quantization of YM theories in the continuum requires the introduction of a gauge fixing which is typically implemented by the so-called Faddeev-Popov (FP) procedure \cite{Faddeev:1967fc}. Such a method provides a very good description of the dynamics of YM theories in the ultraviolet (UV), where perturbation theory is safely applicable. Nevertheless, as pointed out in \cite{Gribov:1977wm}, the assumptions that underpin the FP procedure are not valid in the regime where the theory becomes strongly coupled. The reason is that there exists different gauge-field configurations which satisfy a given gauge condition and are related by a gauge transformation (for sufficiently large values of the product between the gauge coupling and the gauge field). This contradicts an important assumption made in the FP procedure, where the gauge-fixing condition is assumed to be ideal, i.e., it selects just one gauge-field representative per gauge orbit. In fact, in \cite{Gribov:1977wm} such a pathology was explicitly identified in the Landau gauge. It turns out that this is not a particular problem of this gauge choice but a general feature which can be traced back to the non-trivial bundle structure of non-Abelian gauge theories, see \cite{Singer:1978dk}. The inability to completely fix the gauge is known as the Gribov problem and the spurious configurations are known as Gribov copies.

In \cite{Gribov:1977wm}, it was proposed to restrict the path integral of YM theories quantized in the Landau gauge to a domain in which a class of Gribov copies were absent. In fact, those are the infinitesimal Gribov copies, i.e., those that are generated by infinitesimal gauge transformations. Remarkably, in the Landau gauge, it is possible to define a region in field space that is free of infinitesimal Gribov copies. Such a region is known as the Gribov region and has important geometrical properties that makes the restriction of the path integral to the Gribov region a consistent procedure. The Gribov region is bounded in every direction in field space and its boundary is known as the (first) Gribov horizon, it is convex and, very importantly, every gauge orbit cross it at least once \cite{DellAntonio:1991mms}. This last property ensures that the restriction of the path integral to the Gribov region does not leave out any configuration which does not have an equivalent inside it. It should be emphasized that, albeit such a region is free of infinitesimal Gribov copies, it still contains copies that are generated by finite gauge transformations, as shown in \cite{vanBaal:1991zw}. The restriction of the path integral to the Gribov region, proposed in \cite{Gribov:1977wm}, was fully worked out in \cite{Zwanziger:1989mf} by means of a different method. In fact, it was proven in \cite{Capri:2012wx} that extending the results of \cite{Gribov:1977wm} (which were worked out at leading order) to all orders, in a perturbative expansion, coincides with the result obtained in \cite{Zwanziger:1989mf}. As showed in \cite{Zwanziger:1989mf}, the restriction of the path integral of gauge-fixed YM theory in the Landau gauge is effectively implemented by a non-local modification to the gauge-fixed YM action. Remarkably, such a non-locality can be cast in local form by means of the introduction of auxiliary fields. The resulting framework proposed in \cite{Zwanziger:1989mf} implements the restriction of the path integral to the Gribov region in a local and renormalizable way, see also \cite{Vandersickel:2012tz}. This local and renormalizable action is known as the Gribov-Zwanziger (GZ) action. Such an action provides a gluon propagator that identically vanishes at zero momentum and an enhanced ghost propagator in the deep infrared. Such a suppression of the gluon propagator ensures that it violates reflection positivity and, thus, prevents a direct interpretation of gluons as physical excitations in the spectrum, being a potential signature of confinement. Nonetheless, lattice data extracted from simulations in very large lattices displayed a non-vanishing gluon propagator at zero momentum and a non-enhanced ghost propagator in the infrared, see, e.g., \cite{Cucchieri:2007md,Cucchieri:2007rg,Sternbeck:2007ug}. 

In \cite{Dudal:2007cw,Dudal:2008sp}, it was pointed out that the GZ action suffers from IR instabilities and further non-perturbative effects must be taken into account. Such effects are due to the formation of gluon condensates and of the localizing auxiliary fields, which acquire their own dynamics, see also \cite{Dudal:2011gd,Dudal:2019ing}. The inclusion of such effects to the GZ action led to the setup known as Refined GZ (RGZ). Such a framework provides a tree-level gluon propagator which can be fitted very well with the data extracted from very large lattices in the IR. The resulting fitting provides a gluon propagator which features positivity violation as well as complex poles see, e.g., \cite{Cucchieri:2011ig,Dudal:2018cli}. More recently, it was shown that one-loop correction to the gluon propagator within the RGZ framework does not destabilize the good properties of the tree-level propagator, see \cite{deBrito:2024ffa}. Radiative corrections to other correlation functions in the RGZ framework were also computed in \cite{Mintz:2017qri,deBrito:2023qfs,Barrios:2024idr} and the results compare well with the available lattice data. Hence, the RGZ scenario is an effective way of restricting the path integral to the Gribov region and, thus, removing infinitesimal Gribov copies, and taking into account further non-perturbative effects in a local and renormalizable way, see \cite{Vandersickel:2011zc}, in the Landau gauge.  At this stage, an important comment is in order: The (R)GZ action breaks BRST symmetry in an explicit but soft way \cite{Zwanziger:1989mf,Dudal:2008sp}. It should be emphasized that BRST symmetry is a very important outcome of the FP procedure, see \cite{Becchi:1975nq,Baulieu:1981sb,Barnich:2000zw}. Being a soft breaking, one ensures that in the deep UV, the standard FP action and BRST invariance are recovered. Since the (R)GZ framework is developed precisely to amend the issues of the FP method, it is not too surprising that the standard BRST transformations do not correspond to a symmetry of the (R)GZ action. Such a BRST breaking was subject to intense research, see, e.g., \cite{Maggiore:1993wq,Baulieu:2008fy,Dudal:2009xh,Sorella:2009vt,Sorella:2010it,Capri:2010hb,Lavrov:2011wb,Serreau:2012cg,Serreau:2013ila,Dudal:2012sb,Pereira:2013aza,Pereira:2014apa,Lavrov:2013boa,Capri:2014bsa,Cucchieri:2014via,Moshin:2014xka,Schaden:2014bea,Schaden:2015uua}. 

One of the biggest challenges to be faced with the BRST breaking lies on the tentative elimination of Gribov copies in gauges different from the Landau gauge\footnote{It should be said that besides the Landau gauge, a first-principles construction of the (R)GZ action is also available in the maximal Abelian gauge, where BRST is also softly broken, see, e.g., \cite{Capri:2005tj,Capri:2006cz,Capri:2008ak,Capri:2008vk,Capri:2010an,Capri:2015pfa}.}. Over the past decade, a reformulation of the (R)GZ action in the Landau gauge that is compatible with BRST invariance was achieved. The key ingredient for such a construction was the introduction of dressed (gauge-invariant) fields, see \cite{Capri:2015ixa} and the subsequent works \cite{Capri:2015nzw,Capri:2016aqq,Capri:2016gut,Pereira:2016fpn,Capri:2017abz,Capri:2017bfd,Capri:2018ijg}. Remarkably, the correlation functions of the BRST-invariant formulation of the (R)GZ framework (in the Landau gauge) of gluons and FP ghost fields are exactly the same as the corresponding correlation functions computed in the standard BRST-broken (R)GZ scenario (in the Landau gauge). Hence, in practice, the main gain of the BRST-invariant formulation in the Landau gauge lies on the fact that it enables the proposition of a BRST quantization of YM theories in gauges that are connected to the Landau gauge, taking into account the elimination of infinitesimal Gribov copies and providing a consistent control of gauge-parameter dependence. In \cite{Capri:2017bfd}, it was shown that the BRST-invariant RGZ action in linear covariant gauges is renormalizable at all orders in perturbation theory by means of the algebraic renormalization framework \cite{Piguet:1995er}. Hence, there exists a local, BRST-invariant and renormalizable framework that removes infinitesimal Gribov copies from the path integral of YM theories quantized in linear covariant gauges and takes into account further non-perturbative effects such as the formation of condensates.

The features described above are mostly focused in four-dimensional Euclidean YM theories. However, as investigated in \cite{Dudal:2008rm}, the elimination of Gribov copies, as well as the formation of condensates, carry over to three dimensions. In particular, the elimination of Gribov copies as worked out in \cite{Gribov:1977wm,Zwanziger:1989mf} is a geometrical problem that can be formulated in $d$ dimensions. As for the formation of condensates, this requires a detailed analysis due to the peculiarities of quantum field theories in $d=2$, see, e.g., \cite{Dudal:2008xd} (we also refer the reader for a discussion in $d>4$, see \cite{Guimaraes:2016okb}). Hence, in $d=3$, the RGZ framework predicts a gluon propagator that is finite at vanishing momentum and a ghost propagator that is not enhanced in the IR. This agrees with lattice data as well, see \cite{Cucchieri:2011ig}. Besides being a computationally less costly environment thanks to its renormalization properties or numerical advantages\footnote{See also \cite{Huber:2016tvc}.}, working in three dimensions can be very insightful for other reasons. It is well-known that, in three dimensions, a mass-like term can be introduced for the gauge fields through the Chern-Simons (CS) term \cite{Witten:1988hf,Delduc:1989ft,Blasi:1990xz,Delduc:1990je,Dunne:1998qy,Zanelli:2012zz}. Without the YM action, the CS action is topological and has been investigated in very different contexts and several remarkable properties have been established, such as its finiteness \cite{Delduc:1990je}. A rich model that is compatible with gauge invariance and can provide non-trivial insights is the combination of the YM action with the CS one. This leads to the so-called Yang-Mills-Chern-Simons (YMCS) action, which has been vastly explored over the past decades \cite{Deser:1981wh,Deser:1982vy,Pisarski:1985yj,Giavarini:1992xz,DelCima:1997pb,DelCima:1998ur,Azevedo:2024cov}. In this case, the theory is not topological, but still contains a mass of topological nature (arising from a parity-violating term) and is finite, see \cite{DelCima:1998ur,Azevedo:2024cov}. Once again, gauge invariance requires the introduction of a gauge-fixing condition which is hampered by the Gribov problem. As explained above, the restriction of the path integral of YMCS theories in the Landau gauge can be worked out in complete analogy to the standard YM case thanks to its purely geometrical nature within field space. Nevertheless, the dynamics in YMCS theories is different from the one of pure YM theories. The gluon propagator is different and, in the case where infinitesimal Gribov copies are eliminated, it contains mass-like parameters of different nature: Those arising from the elimination of Gribov copies and the one which is present thanks to the CS term. In \cite{Canfora:2013zza,Ferreira:2020kqv,Ferreira:2021gqh,Felix:2021eoq}, a systematic investigation about the elimination of infinitesimal Gribov copies in YMCS theories in the Landau, maximal Abelian and linear covariant gauges was carried out. Moreover, in \cite{Gomez:2015aba}, the inclusion of Higgs fields was also explored. Such a system has a rich tree-level gluon propagator that, thanks to the competition of different mass parameters, provides a rich phase diagram, leading to the possibility of characterizing would-be confined and deconfined phases. This is a suitable scenario to investigate how the pole-structure engendered by the elimination of Gribov copies coexists with other mass parameters as the CS mass or the Higgs vacuum expectation value in a reasonably simple setting. 

One remarkable property concerning the RGZ action in four dimensions is that the inclusion of all terms that are necessary to eliminate infinitesimal Gribov copies and account for the formation of dimension-two condensates does not affect the renormalization properties of standard YM theories gauge-fixed by the FP procedure \cite{Dudal:2008sp,Capri:2017bfd}. In the context of YMCS theories, a natural question arises: Does the elimination of infinitesimal Gribov copies affect the finiteness of such a theory? In other words, is the Refined Gribov-Zwanziger Yang-Mills-Chern-Simons (RGZ-YMCS) theory finite? The goal of this paper is to answer this question and provide an algebraic proof that the theory remains finite in linear covariant gauges. The proof follows the algebraic renormalization framework \cite{Piguet:1995er} and uses results from the YMCS literature \cite{DelCima:1998ur}, as well as inspired techniques developed in \cite{Capri:2017bfd} for the proof of renormalizability of the RGZ in linear covariant gauges in four dimensions. It is important to emphasize that the elimination of Gribov copies in harmony with BRST invariance requires the introduction of a non-polynomial field, the gauge-invariant dressed field, a fact that makes the algebraic analysis much more subtle. 

This paper is organized as follows: In Sect.~\ref{Sect:RGZYMCS} we provide a short review of the elimination of infinitesimal Gribov copies in YMCS theories quantized in linear covariant gauges. In Sect.~\ref{ARSS}, we cast the RGZ-YMCS action in a suitable form for the use of the standard techniques employed in the Algebraic Renormalization program. We also introduce an extended BRST symmetry and list the ensuing Ward identities that will play a pivotal role in this work. Sect.~\ref{FYMCSRGZ} contains the proof that RGZ-YMCS theories are finite in linear covariant gauges. Finally, we collect our conclusions and perspectives.

\section{The elimination of Gribov copies in Yang-Mills-Chern-Simons theories in linear covariant gauges \label{Sect:RGZYMCS}} 

\subsection{Setting conventions: YMCS action in linear covariant gauges\label{Sub.Sect:RGZYMCS.Conv}} 

The YMCS action defined in three Euclidean dimensions and with $SU(N)$ gauge group is defined as
\begin{equation}
	S_{\rm YMCS} = \frac{1}{4}\int {\rm d}^3x~ F_{\mu\nu}^a F_{\mu\nu}^a  -iM\int {\rm d}^3x~ \epsilon_{\mu\rho\nu}\left(\frac{1}{2}A_\mu^a \partial_\rho A_\nu^a + \frac{g}{3!}f^{abc} A_\mu^a A_\rho^b A_\nu^c\right)\,,
\label{conv.1}
\end{equation}
with $F_{\mu\nu}^a = \partial_\mu A^a_\nu - \partial_\nu A^a_\mu +gf^{abc}A^b_\mu A^c_\nu$ being the field strength. The gauge coupling is denoted by $g$ and has mass-dimension one-half. The parameter $M$ is the CS mass, $f^{abc}$ are the structure constants of the $SU(N)$ gauge group and $\epsilon_{\mu\rho\nu}$ is the totally anti-symmetric Levi-Civita symbol. The action \eqref{conv.1} is invariant under infinitesimal gauge transformations and its standard quantization requires a gauge-fixing procedure. This can be achieved by the FP procedure or, more directly, through the BRST quantization. Our focus is the quantization of \eqref{conv.1} in linear covariant gauges. This is achived by adding to \eqref{conv.1} a BRST-exact term as follows,
\begin{equation}\label{slcg}
	S = S_{\rm YMCS} + s\int {\rm d}^3x~ \bar{c}^a\left(\partial_\mu A_\mu^a - \frac{\alpha}{2} b^a \right) = S_{\rm YMCS} + \int {\rm d}^3x\left(b^a\partial_\mu A^a_\mu - \frac{\alpha}{2}b^a b^a+ \bar{c}^a \partial_\mu D_\mu^{ab}c^b \right),
\end{equation}
where $c^a,\bar{c}^a,b^a$ are the FP ghost, antighost and the Nakanishi-Lautrup fields, respectively. The parameter $\alpha$ is a non-negative gauge parameter and the limit $\alpha\to 0$ leads to the Landau gauge. The BRST operator $s$ has Grassmann number one and is nilpotent, i.e., $s^2 = 0$. The elementary fields transform as
\begin{align}
s A^a_\mu &= - D^{ab}_\mu c^b\,,  &&s c^a = \frac{g}{2}f^{abc}c^b c^c\,, \nonumber \\ 
s \bar{c}^a &= b^a\,, && s b^a = 0\,.
\label{Eq:BRST.ElementaryFields}
\end{align}

The action $S$ can be quantized by means of path-integral methods leading to 
\begin{equation}
	\EuScript{Z}_{\rm YMCS} = \int [\EuScript{D}\mu]_{\rm YM}~{\rm e}^{-S}\,,
	\label{Zymcs}
\end{equation}
with 
\begin{equation}
[\EuScript{D}\mu]_{\rm YM} = [\EuScript{D}A][\EuScript{D}b][\EuScript{D}c][\EuScript{D}\bar{c}]\,.
\end{equation}
The path-integral \eqref{Zymcs} is plagued by Gribov copies \cite{Ferreira:2021gqh} and a naive integration should render an overcounting of physically equivalent configurations even after the gauge fixing. In the next subsections, we shall summarize how the elimination of infinitesimal Gribov copies is achieved in this case and what are the underlying modifications to the starting point action.

\subsection{Elimination of infinitesimal Gribov copies: A short overview\label{Sub.Sect:RGZYMCS.Elim.GribovCopies}} 

In this subsection, we will shortly review the elimination of (infinitesimal) Gribov copies in YMCS theories with $SU(N)$ gauge group quantized in linear covariant gauges. A detailed construction of the underlying action that implements such an elimination can be found in \cite{Ferreira:2021gqh} and references therein. In fact, the elimination of (infinitesimal) Gribov copies in standard Euclidean four-dimensional YM theories has a reasonably long body of literature in linear covariant gauges, see, e.g.,  \cite{Sobreiro:2005vn,Capri:2015pja,Capri:2015ixa,Capri:2015nzw,Capri:2016aqq,Capri:2016gut,Capri:2016aif,Capri:2017bfd,Mintz:2018hhx,Capri:2018ijg}. The great challenge in moving away from the Landau gauge (where the elimination of infinitesimal Gribov copies was reasonably well-understood in the standard (R)GZ framework) lies in the fact that the FP operator in linear covariant gauges is not Hermitian. Keeping in mind that normalizable zero modes of such an operator correspond to infinitesimal Gribov copies, one has to ensure that the functional integral is restricted to a region free of those configurations. In the Landau gauge, hermiticity guarantees a real spectrum and thus it is meaningful to integrate over the region where the FP operator is positive. This property is lost in linear covariant gauges, making a systematic elimination of (infinitesimal) Gribov copies much harder. Moreover, the soft BRST-breaking in the Landau gauge hampered a would-be BRST quantization consistent with the elimination of gauge copies. Next to that, since linear covariant gauges carry a gauge parameter, BRST invariance becomes essential to ensure that correlation functions of gauge-invariant quantities are, indeed, gauge-parameter independent. In the series of works mentioned above, the key ingredient to give a step forward towards the elimination of Gribov copies in linear covariant gauges was the introduction of a gauge-invariant field $A^h_\mu \equiv A^{h,a}_\mu T^a$ (with $\left\{T^a\right\}$ being the generators of $SU(N)$) which might be viewed as a ``dressed'' gauge field, see, e.g., \cite{Zwanziger:1990tn,Lavelle:1995ty}. We refer the reader to, e.g., the Appendix of \cite{Capri:2015ixa} for a step-by-step construction of $A^h_\mu$. The gauge invariance of $A^h_\mu$ has enabled for a reformulation of the RGZ action in the Landau gauge that is compatible with BRST symmetry, although predicting exactly the same correlation functions involving gauge and FP ghost fields. Nevertheless, this opens up the possibility to propose a BRST quantization which incorporates the elimination of Gribov copies in linear covariant gauges and it is even possible to provide a geometrical interpretation to the resulting action by a restriction of the path integral to a specific domain, which is not associated directly with the FP operator in linear covariant gauges, but to a ``dressed'' FP operator, see, e.g., \cite{Capri:2015ixa}. 

The path integral that we are interested on is defined by
\begin{equation}
	\EuScript{Z}^{\rm RGZ}_{\rm YMCS} = \int [\EuScript{D}\mu]_{\rm GZ}~{\rm e}^{-S^{\rm RGZ}_{\rm YMCS}}\,,
	\label{ZRGZYMCS}
\end{equation}
with 
\begin{equation}
[\EuScript{D}\mu]_{\rm GZ} = [\EuScript{D}\mu]_{\rm YM}[\EuScript{D}\varphi][\EuScript{D}\bar{\varphi}][\EuScript{D}\omega][\EuScript{D}\bar{\omega}][\EuScript{D}\xi][\EuScript{D}\tau][\EuScript{D}\eta][\EuScript{D}\bar{\eta}]\,,
\label{RGZmeasure}
\end{equation}
and
\begin{equation}\label{rgz2}
	\begin{split}
		S^{\rm RGZ}_{\rm YMCS } &= S - \int {\rm d}^3x~\Big(\bar{\varphi}_\mu^{ac}\mathcal{M}^{ab}(A^h){\varphi}_\mu^{bc}-\bar{\omega}_\mu^{ac}\mathcal{M}^{ab}(A^h){\omega}_\mu^{bc}\Big) + \gamma^2 \int {\rm d}^3x~g f^{abc}A_\mu^{h,a} (\varphi_\mu^{bc}+\bar{\varphi}_\mu^{bc})\\ 
		&+ \frac{m^2}{2}\int {\rm d}^3x~ A^{h,a}_\mu A^{h,a}_\mu - \mu^2 \int {\rm d}^3x~ (\bar{\varphi}^{ab}_\mu \varphi^{ab}_\mu - \bar{\omega}^{ab}_\mu  \omega^{ab}_\mu ) + \int {\rm d}^3x~\Big(\tau^a\partial_\mu A_\mu^{h,a}-\bar{\eta}^a\mathcal{M}^{ab}(A^h)\eta^b\Big)\,.
	\end{split}
\end{equation}
As an important remark, the so-called GZ action (which in the context of YMCS theories will be called GZ-YMCS action) is the same as the one in \eqref{rgz2} apart from the addition of the so-called refining condensates. It is given by 
\begin{equation}\label{gz2}
	\begin{split}
		S^{\rm GZ}_{\rm YMCS } &= S - \int {\rm d}^3x~\Big(\bar{\varphi}_\mu^{ac}\mathcal{M}^{ab}(A^h){\varphi}_\mu^{bc}-\bar{\omega}_\mu^{ac}\mathcal{M}^{ab}(A^h){\omega}_\mu^{bc}\Big) + \gamma^2 \int {\rm d}^3x~g f^{abc}A_\mu^{h,a} (\varphi_\mu^{bc}+\bar{\varphi}_\mu^{bc})\\ 
		&+ \int {\rm d}^3x~\Big(\tau^a\partial_\mu A_\mu^{h,a}-\bar{\eta}^a\mathcal{M}^{ab}(A^h)\eta^b\Big)\,.
	\end{split}
\end{equation}
Several comments are in order: The parameters $(\gamma^2, m^2,\mu^2)$ are, respectively, the Gribov parameter and the refining mass parameters that are introduced self-consistently along with the elimination of infinitesimal Gribov copies. They are self-consistent in the sense that they are not free, but fixed by their corresponding gap equations, see \cite{Dudal:2011gd,Dudal:2019ing}. In particular, in the Landau gauge, it is clear that $\gamma^2$ arises due to the boundedness of the Gribov region in this gauge. Yet, in linear covariant gauges, one can associate the elimination of Gribov copies to the region defined by the field configurations that satisfy the linear covariant gauge condition and are contained in $-\partial_\mu D^{ab}_\mu (A^h) > 0$, where the covariant derivative is defined with respect to the dressed field $A^h$. The other mass parameters, i.e., $(m^2,\mu^2)$ are introduced due to infrared instabilities of the GZ action as proposed in \cite{Dudal:2007cw,Dudal:2008sp}. The RGZ-YMCS action has a large set of extra fields $(\bar{\varphi},\varphi,\bar{\omega},\omega, \xi,\tau, \bar{\eta},\eta)$ with respect to the standard FP action and they are introduced in order to localize the RGZ action, in the context of YMCS theories in linear covariant gauges. We have introduced the notation $\mathcal{M}^{ab}(A^h) \equiv -\partial_\mu D^{ab}_\mu (A^h) > 0$, which is the analogue of the FP operator but with the dressed field\footnote{Since the field $A^h$ is transverse, the operator $\mathcal{M}(A^h)$ is Hermitian.}. One can integrate out the auxiliary fields and obtain a non-local action with two sources of non-localities: One that arises from the so-called Horizon function, i.e., the term that implements the restriction of the integration domain to a region free of infinitesimal copies. Its non-locality is due to the presence of the inverse of the operator $\mathcal{M}^{ab}(A^h)$ (which is well-defined inside such a region since it is free of its zero modes). The other non-locality arises from the dressed field $A^h_\mu$ itself, which can be expressed as an infinite series of the gauge field containing inverse Laplacians. We refer the interested reader to \cite{Capri:2017bfd} for more details of such localization procedure. We emphasize that the localization of the dressed field $A^h_\mu$ requires the introduction of a Stueckelberg-like field $\xi^a$, giving rise to a non-polynomial, albeit local, expression for $A^h_\mu$. Another very important property of \eqref{rgz2} is that, if $\alpha$ is set to zero, one recovers the standard RGZ action in the Landau gauge (which can be constructed by a direct restriction of the path integral to the so-called Gribov region), thanks to the decoupling of the Stueckelberg-like field. Finally, the action \eqref{rgz2} is invariant under a set of BRST transformations that is nilpotent, i.e., 
\begin{equation}\label{brst}
	\begin{aligned}
		&sA^a_\mu = -D_\mu^{ab}c^b, & &sc^a = \frac{g}{2}f^{abc}c^b c^c,\\
		&s\bar{c}^a = b^a, & &sb^a = 0,\\
		&s\bar{\omega}_\mu^{ab} = 0, & &s\bar{\varphi}_\mu^{ab} = 0,\\
		&s\varphi_\mu^{ab} =0, & &s{\omega}_\mu^{ab} = 0,\\
		&sA_\mu^{h,a}=0, & &s\tau^a =0,\\
		&s\bar{\eta}^a=0, & &s\eta^a=0,\\
		&sh^{ij} = -igc^a(T^a)^{ik}h^{kj}, & &s\xi^a = g^{ab}(\xi)c^b,
	\end{aligned}
\end{equation}
with
\begin{equation}\label{sxi}
	g^{ab}(\xi) = -\delta^{ab} + \frac{g}{2}f^{abc}\xi^c -\frac{g^2}{12}f^{amr}f^{mbq}\xi^q\xi^r + O(\xi^3)\,,
\end{equation}
and
\begin{equation}
	A_\mu^{h,a}T^a = h^\dagger A_\mu h +\frac{i}{g}h^\dagger\partial_\mu h\,,\quad {\rm and} \quad h = e^{ig\xi^aT^a}\equiv e^{ig\xi}\,.
\end{equation}
The field $\xi^a$ is precisely the Stueckelberg-like field that is necessary to be introduced in order to localize $A^h_\mu$ and does not appear explicitly in the writing of \eqref{rgz2}. Thus, the action \eqref{rgz2} corresponds to a local (but non-polynomial) framework which eliminates infinitesimal Gribov copies of YMCS theories quantized in linear covariant gauges with gauge group $SU(N)$. The main goal of this work is to prove that such an action leads to a finite theory, i.e., dealing with the elimination of infinitesimal Gribov copies and further instabilities generated by such a removal does not affect the well-known properties of the YMCS theories quantized in the FP framework.

\section{Algebraic Renormalization of the RGZ-YMCS Action: Setting the Stage \label{ARSS}}

In this Section, we will endow the action $S^{\rm RGZ}_{\rm YMCS}$ with all the relevant structure that facilitates the use of the Algebraic Renormalization framework. Moreover, we will make use of some of the techniques developed in the analysis performed in \cite{Capri:2017bfd} and \cite{Capri:2016ovw}, where the non-polynomial structure of the $A^h_\mu$ field is present with and without the terms that eliminate Gribov copies in linear covariant gauges. We should also stress that, unlike \cite{Capri:2017bfd} and \cite{Capri:2016ovw}, we are dealing with a theory in three Euclidean dimensions in this work. As such, due to the fact that the gauge-coupling $g^2$ has canonical mass dimension one, the power-counting is severely affected and we are dealing with a superrenormalizable theory. In fact, previous studies involving the Algebraic Renormalization of pure YM theories supplemented with parity-preserving mass terms in three-dimensions have led to the conclusion that those theories are finite, see, e.g., \cite{Dudal:2004ch,Dudal:2006ip,Dudal:2008zb}. From a structural point of view, this work takes into account parity-preserving and violating mass terms in harmony with gauge invariance. Aftermath, we conclude that the inclusion of all of these massive structure together with the elimination of infinitesimal Gribov copies still renders a finite theory. In order to proceed with the use of the Algebraic Renormalization setup as in \cite{Capri:2017bfd}, we have to introduce, besides the usual external sources coupled to the non-linear BRST transformations, a set of sources that takes (specific) physical values at the end of the renormalization procedure. We shall see how the different power-counting will affect the characterization of the counterterm.

\subsection{The introduction of external sources \label{SubSec.Ext.Sources}}

In order to accommodate the non-linear nature of the BRST transformations \eqref{brst}, we add to the original action a set of sources coupled to the non-linear BRST transformations. On top of that, due to the composite nature of $A^{h,a}_\mu$, we also couple it to an external source for future use. Thus, we add to $S^{\rm RGZ}_{\rm YMCS}$ the following term,
\begin{equation}
	\Sigma_{\rm ext} = \int {\rm d}^3x~ \Big(-\Omega^a_\mu D_\mu^{ab}c^b + \frac{g}{2}L^af^{abc}c^b c^c  + K^a g^{ab}(\xi)c^b + \mathcal{J}^a_\mu A^{h,a}_\mu\Big)\,.
	\label{BRSTsources}
\end{equation}
The sources $(\Omega^a_\mu, L^a, K^a)$ are coupled to the BRST transformations of the fields $(A^a_\mu, c^a,\xi^a)$ and the source $\mathcal{J}^a_\mu$ is directly coupled to $A^{h,a}_\mu$. Since those sources are coupled to BRST variations and the BRST operator is nilpotent, in order to preserve the BRST-invariance of $\Sigma_{\rm ext}$, we have to demand that those sources transform trivially under the action of the BRST operator, i.e., $s\Omega^a_\mu = sL^a = s K^a = s\mathcal{J}^a_\mu = 0$.

Due to the presence of the condensates $\langle A^{h,a}_\mu (x) A^{h,a}_\mu (x) \rangle$ and $\langle \bar{\varphi}^{ab}_\mu (x) \varphi^{ab}_\mu (x) - \bar{\omega}^{ab}_\mu (x) \omega^{ab}_\mu (x)\rangle$ in the RGZ-YMCS action, we introduce appropriate sources coupled to such local composite operators, following \cite{Knecht:2001cc,Verschelde:2001ia},
\begin{equation}
	\Sigma_{\rm cond} = \int {\rm d}^3x~ \Big[J(A^{h,a}_\mu A^{h,a}_\mu) - \rho(\bar{\varphi}^{ab}_\mu \varphi^{ab}_\mu - \bar{\omega}^{ab}_\mu  \omega^{ab}_\mu ) \Big]\,.
	\label{CondAction}
\end{equation}
Some remarks are in order: The composite operators in \eqref{CondAction} are BRST invariant and therefore so are the sources coupled to them, i.e., $sJ = s\rho = 0$. Such a term will be added in replacement to the condensate terms that are present in the RGZ-YMCS action, i.e., the first two terms of the second line of \eqref{rgz2}. In the end, those sources should attain their physical values, which are, in the present case,
\begin{equation}
J\,\Big|_{\rm phys} = \frac{m^2}{2}, \qquad \rho\,\Big|_{\rm phys} = \mu^2.
	\label{PhysValuesCond}
\end{equation}
In general, when introducing (dimension-two) local composite operators in four dimensions, one also needs to introduce source-squared terms to account for divergences arising from $\langle \EuScript{O}(x)\EuScript{O}(y)\rangle_{x\to y} $, with $\EuScript{O}(x)$ being the local composite operator. It is well understood that for the local composite operators of auxiliary fields, such a term is absent thanks to remarkable properties of the localizing-sector of the RGZ action \cite{Dudal:2008sp}. As for the condensate arising from $\langle A^{h,a}_\mu (x) A^{h,a}_\mu (x) \rangle$, a source-squared term should be added, in principle. However, since we are in three dimensions, such a term is not allowed by power-counting (the canonical dimensions of fields and sources are collected in Tables~\ref{Table 1} and \ref{Table 2}). Yet one could still introduce terms of the sort $\zeta_1 J$ and $\zeta_2 \rho$, with $\zeta_1$ and $\zeta_2$ being dimension-one parameters. In practice, those terms are harmless, since they are just pure additive field/source-independent contributions after the physical limit is taken and thus they do not contribute to correlation functions.

We now focus on the introduction of a different class of sources. In the original BRST-softly broken (R)GZ framework in the Landau gauge \cite{Zwanziger:1989mf,Dudal:2008sp}, a set of sources was introduced in order to restore BRST invariance. Such a procedure allows for the use of the powerful cohomology techniques that are present in BRST-invariant theories. In the end of the renormalization analysis, the sources are tuned to their physical value, and BRST returns to be a softly broken symmetry. The renormalization properties of those different theories (i.e., in the presence of the sources and at their physical values) are the same when the symmetry breaking is soft, see \cite{Symanzik:1969ek}. In the present case, BRST invariance is present from the beginning. Nevertheless, the introduction of such sources remains useful, as done in \cite{Capri:2017bfd}. Those sources are introduced as follows,
\begin{equation}
	\begin{split}
		S_{\gamma^2} \equiv \int {\rm d}^3x~\gamma^2 g f^{abc}A^{h,a}_\mu(\varphi+\bar{\varphi})^{bc}_\mu \quad \longrightarrow \quad &\Sigma_{\gamma^2} \equiv \int {\rm d}^3x~ \Big[M^{ai}_\mu D^{ab}_\mu(A^h)\varphi^{bi} + V^{ai}_\mu D^{ab}_\mu(A^h)\bar{\varphi}^{bi} + N^{ai}_\mu D^{ab}_\mu(A^h)\omega^{bi}\\
		&+ U^{ai}_\mu D^{ab}_\mu(A^h)\bar{\omega}^{bi} - M^{ai}_\mu V^{ai}_\mu + N^{ai}_\mu U^{ai}_\mu\Big]\,,
	\end{split}
	\label{sourcesGZ}
\end{equation}
where we employed the multi-index notation $i=(a,\mu)$. We have introduced the sources $(M,V,N,U)^{ab}_{\mu\nu}\equiv (M,V,N,U)^{ai}_\mu$. The multi-index contains a color and a (space)time index. We should emphasize that we introduce the terms as in \eqref{sourcesGZ} inspired by the way that sources are introduced in the standard (R)GZ action in the Landau gauge, see \cite{Zwanziger:1989mf,Dudal:2008sp}. Rewriting this part of the full action in this way gives us important Ward identities that we shall explore later on. Moreover, the quadratic terms on the sources are allowed by power-counting. Once again, we fix their coefficients based on what is already known in the standard Landau gauge construction. The introduction of the sources enlarges the original action $S_{\gamma^2}$ that must be recovered under an appropriate physical limit of the sources, i.e.,  
\begin{equation}\label{sourcesgrib}
		M^{ab}_{\mu\nu}|_{\rm phys} = V^{ab}_{\mu\nu}|_{\rm phys} = \gamma^2 \delta^{ab}\delta_{\mu\nu}\,,\quad {\rm and} \quad
		N^{ab}_{\mu\nu}|_{\rm phys} = U^{ab}_{\mu\nu}|_{\rm phys} = 0\,.
\end{equation}
The sources are BRST-singlets, so that BRST invariance is preserved,
\begin{equation}
	sM^{ab}_{\mu\nu} = sV^{ab}_{\mu\nu} = sN^{ab}_{\mu\nu} = sU^{ab}_{\mu\nu} = 0\,.
\end{equation}
The presence of these sources will be convenient for the construction of the counterterm action later. 

There are other non-linear terms that show to be better controlled by the introduction of extra sources, as pointed out in \cite{Capri:2017bfd}. We repeat the procedure here and add the following term,
\begin{equation}
	\Sigma_{\rm extra} = \int {\rm d}^3x~ [-\Xi^a_\mu D^{ab}(A^h)\eta^b + X^i \eta^a\bar{\omega}^{ai} + Y^i \eta^a \bar{\varphi}^{ai} + \bar{X}^{abi}\eta^a\omega^{bi} + \bar{Y}^{abi}\eta^a \varphi^{bi}]\,.
\end{equation}
As usual, the new sources are BRST singlets, namely,
\begin{equation}
	 s\,\Xi^a_\mu = sX^i = sY^i = s\bar{X}^{abi} = s\bar{Y}^{abi} = 0\,.
\end{equation}
After the introduction of this whole set of external sources, we are ready to write down the complete action which will be our target in the algebraic renormalizability analysis. The action reads
\begin{eqnarray}
\Sigma &=& S - \int {\rm d}^3x~\Big(\bar{\varphi}_\mu^{ac}\mathcal{M}^{ab}(A^h){\varphi}_\mu^{bc}-\bar{\omega}_\mu^{ac}\mathcal{M}^{ab}(A^h){\omega}_\mu^{bc}\Big)+ \int {\rm d}^3x~\Big(\tau^a\partial_\mu A_\mu^{h,a}-\bar{\eta}^a\mathcal{M}^{ab}(A^h)\eta^b\Big)+\Sigma_{\gamma^2}\nonumber\\
&+& \Sigma_{\rm cond} + \Sigma_{\rm ext} + \Sigma_{\rm extra}\,.
\label{Sigma.1}
\end{eqnarray}
and it enjoys BRST invariance, i.e., 
\begin{equation}
	s\Sigma = 0\,.
\end{equation}
Also, we see that, assuming the external sources physical values to be zero
\begin{equation}
	\Omega^a_\mu|_{\rm phys}  = L^a|_{\rm phys}  = \mathcal{J}^a_\mu|_{\rm phys}  = K^a|_{\rm phys}  = \Xi^a_\mu|_{\rm phys}  = X^i|_{\rm phys}  =  Y^i|_{\rm phys}  = \bar{X}^{abi}|_{\rm phys}  = \bar{Y}^{abi}|_{\rm phys}  = 0\,,
\end{equation}
together with equations \eqref{sourcesgrib} and \eqref{PhysValuesCond}, the total action \eqref{Sigma.1} reduces to the RGZ-YMCS action \eqref{rgz2},
\begin{equation}
	\Sigma\,\Big|_{\rm phys} = S^{\rm RGZ}_{\rm YMCS}\,.
\end{equation}
From now on, we will work out the algebraic renormalization of $\Sigma$. Before that, however, it is useful to define an extended BRST symmetry $Q$, which will significantly simplify the analysis. 

\subsection{Extented BRST symmetry}
As introduced in \cite{Piguet:1984js} and explored in \cite{Capri:2016gut,Capri:2017bfd}, it is useful to introduce a BRST transformation for the gauge parameter $\alpha$, i.e., introduce it as a BRST-doublet. 
\begin{equation}
	s\alpha = \chi\, , \quad {\rm and} \quad s\chi = 0\,,
\end{equation}
where $\chi$ is a Grassmannian parameter of ghost number 1. At the end of the algebraic analysis, we can set $\chi$ to zero. This is a very efficient way of controlling $\alpha$-dependence of correlation functions by means of functional identities. Being introduced in a BRST-doublet, the gauge parameter can only enter in the trivial part of the cohomology of the BRST operator. The action $\Sigma$, given by equation \eqref{Sigma.1}, obeys another exact symmetry, defined by the transformations below,
\begin{equation}
	\delta\Sigma =0\, ,
\end{equation}
with Grassmmannian operator $\delta$ acting on the fields and sources as
\begin{equation}\label{delta}
	\begin{aligned}
		&\delta \varphi^{ai} = \omega^{ai}\,, & &\delta \omega^{ai} = 0\,,\\
		&\delta\bar{\omega}^{ai} = \bar{\varphi}^{ai}\,, & &\delta\bar{\varphi}^{ai} = 0\,,\\
		&\delta N^{ai}_\mu = M^{ai}_\mu\,, & &\delta M^{ai}_\mu = 0\,,\\
		&\delta V^{ai}_\mu = U^{ai}_\mu\,, & &\delta U^{ai}_\mu = 0\,,\\
		&\delta Y^i = X^i\,, & &\delta X^i = 0\,,\\
		&\delta \bar{X}^{abi} = -\bar{Y}^{abi}\,, & &\delta\bar{Y}^{abi} = 0\,,
	\end{aligned}
\end{equation}
which clearly shows that $\delta^2=0$, that is, it is a nilpotent operator. Since the operators $s$ and $\delta$ anticommute, i.e., $\{s,\delta\}=0$, we can combine them into an extended, nilpotent operator $Q$, such that
\begin{equation}
	Q = s + \delta\,, \qquad Q^2 = 0\,.
\end{equation}
The transformations of fields and sources under $Q$ are given by
\begin{equation}\label{Qtransf.1}
	\begin{aligned}
		&QA^a_\mu = -D_\mu^{ab}c^b\,, & &Qc^a = \frac{g}{2}f^{abc}c^b c^c\,,\\
		&Q\bar{c}^a = b^a\,, & &Qb^a = 0\,,\\
		&Q \varphi^{ai} = \omega^{ai}\,, & &Q \omega^{ai} = 0\,,\\
		&Q\bar{\omega}^{ai} = \bar{\varphi}^{ai}\,, & &Q\bar{\varphi}^{ai} = 0\,,\\
		&QA_\mu^{h,a}=0\,, & &Q\tau^a =0\,,\\
		&Q\bar{\eta}^a=0\,, & &Q\eta^a=0\,,\\
		&Q\xi^a = g^{ab}(\xi)c^b\,, & &Q\alpha = \chi\,,\\
		&Q\chi = 0\,, & &Q N^{ai}_\mu = M^{ai}_\mu\,,\\
		&Q M^{ai}_\mu = 0\,, & &Q V^{ai}_\mu = U^{ai}_\mu\,,\\
		&Q U^{ai}_\mu = 0\,, & &Q Y^i = X^i\,,\\
		&Q X^i = 0\,, &	&Q \bar{X}^{abi} = -\bar{Y}^{abi}\,,\\
		&Q\bar{Y}^{abi} = 0\,, &	&Q\Omega^a_\mu = 0\,,\\
		&QL^a = 0\,, & &Q\mathcal{J}^a_\mu = 0\,,\\
		&QK^a = 0\,, & &QJ = 0\,,\\
		&Q\Xi^a_\mu = 0\,.
	\end{aligned}
\end{equation}
A comment is in order: The auxiliary localizing fields $(\bar{\varphi},\varphi)^{ai}$ and $(\bar{\omega},\omega)^{ai}$ behave as doublets under $Q$-transformations. In this sense, those fields only appear in the trivial part of the $Q$-cohomology. As such, the source that is used to introduce the refining condensate of localizing fields can be introduced as a $Q$-doublet by introducing the following transformations,
\begin{equation}\label{Qtransf.2}
Q\rho = \sigma\,\quad {\rm and } \quad Q\sigma=0\,.
\end{equation}
where we introduced a source $\sigma$, in order to form a $Q$-doublet. 
The action $\Sigma$, compatible with the extended BRST operator $Q$ with the transformations defined by \eqref{Qtransf.1} and \eqref{Qtransf.2}, is finally written as
\begin{equation}\label{modtotact}
	\begin{split}
		\Sigma &= S_{\rm YMCS} + \int {\rm d}^3x \left(b^a\partial_\mu A_\mu^a - \frac{\alpha}{2}b^a b^a -\frac{1}{2}\chi\bar{c}^ab^a + \bar{c}^a \partial_\mu D_\mu^{ab}c^b \right) 
		- \int {\rm d}^3x~\Big(\bar{\varphi}_\mu^{ac}\mathcal{M}^{ab}(A^h){\varphi}_\mu^{bc}-\bar{\omega}_\mu^{ac}\mathcal{M}^{ab}(A^h){\omega}_\mu^{bc}]\Big)\\
		&+ \int {\rm d}^3x~(\tau^a\partial_\mu A_\mu^{h,a}-\bar{\eta}^a\mathcal{M}^{ab}(A^h)\eta^b) + \int {\rm d}^3x~ \Big[J(A^{h,a}_\mu A^{h,a}_\mu) - \rho(\bar{\varphi}^{ab}_\mu \varphi^{ab}_\mu - \bar{\omega}^{ab}_\mu  \omega^{ab}_\mu ) -\sigma\bar{\omega}^{ab}_\mu\varphi^{ab}_\mu \Big]+ \Sigma_{\gamma^2} + \Sigma_{\rm ext}  \\ 
		& + \Sigma_{\rm extra}\,,
	\end{split}
\end{equation}
with
\begin{equation}
	Q\Sigma = 0\,.
\end{equation}
The action (\ref{modtotact}) is reduces to the action (\ref{rgz2}) when the sources assume the aforementioned physical values (apart from vacuum field-independent contributions) and together with the condition $\chi= 0$.

This extended symmetry significantly improves the algebraic characterization of the most general counterterm, since the $\delta$-component of the $Q$-transformations introduces a large set of $Q$-doublets. Consequentely, those $Q$-doublets are in the trivial part of the cohomology of $Q$, avoiding a proliferation of possible candidate terms in the non-trivial part of the cohomology. As usual in the Algebraic Renormalization setup, we characterize the most general counterterm which respects the conservation of ghost-number, locality and has the appropriate dimensionality. In fact, the use of the multi-index $i=(a,\mu)$ introduces another quantum number that we call $U(f)$ charge. Moreover, the ghost-like localizing fields $(\bar{\eta},\eta)$ carry a ghost number that is independent from the standard FP ghost number. Therefore, we assign different names to make this clear, i.e., $FP$-ghost number for the standard ghost number and $\eta$-ghost number for the quantum number associated with the new localizing fields. The quantum numbers of fields, sources and parameters of the theory are collected in Tables \ref{Table 1} and \ref{Table 2}. In the next subsection, we list the set of functional identities that allows us to algebraically characterize the most general counterterm that has to be added in the renormalization procedure. One of the greatest advantages of the algebraic renormalization framework is its blindness to regularization schemes, which is a very welcome feature in the presence of the Levi-Civita symbol.

\begin{table}
	\begin{tabular}{| c | c | c | c | c | c | c | c | c | c | c | c | c | c | c |}
		\hline
		Fields & $A$ & $b$ & $c$ & $\bar{c}$ & $\xi$ & $\bar{\varphi}$ & $\varphi$ & $\bar{\omega}$ & $\omega$ & $\alpha$ & $\chi$ & $\tau$ & $\eta$ & $\bar{\eta}$ \\
		\hline
		Dimension & 1/2 & 3/2 & -1/2 & 3/2 & -1/2 & 1/2 & 1/2 & 1/2 & 1/2 & 0 & 0 & 3/2 & 1/2 & 1/2 \\
		\hline
		FP-ghost number & 0 & 0 & 1 & -1 & 0 & 0 & 0 & -1 & 1 & 0 & 1 & 0 & 0 & 0 \\
		\hline
		$\eta$-ghost number & 0 & 0 & 0 & 0 & 0 & 0 & 0 & 0 & 0 & 0 & 0 & 0 & 1 & -1 \\
		\hline
		$U(f)$ charge & 0 & 0 & 0 & 0 & 0 & -1 & 1 & -1 & 1 & 0 & 0 & 0 & 0 & 0 \\
		\hline
		Nature & B & B & F & F & B & B & B & F & F & B & F & B & F & F \\
		\hline
	\end{tabular}
	\caption{The quantum numbers of fields and their nature, i.e., bosonic (B) meaning that they are commuting and fermionic (F) that implies an anti-commuting field.}
	\label{Table 1}
\end{table}
\begin{table}
	\begin{tabular}{| c | c | c | c | c | c | c | c | c | c | c | c | c | c | c | c | c |}
		\hline
		Sources & $\Omega$ & $L$ & $K$ & $\mathcal{J}$ & $M$ & $N$ & $U$ & $V$ & $J$ & $\rho$ & $\sigma$ & $\Xi$ & $X$ & $Y$ & $\bar{X}$ & $\bar{Y}$ \\
		\hline
		Dimension & 5/2 & 7/2 & 7/2 & 5/2 & 3/2 & 3/2 & 3/2 & 3/2 & 2 & 2 & 2 & 3/2 & 2 & 2 & 2 & 2 \\
		\hline
		FP-ghost number & -1 & -2 & -1 & 0 & 0 & -1 & 1 & 0 & 0 & 0 & 1 & 0 & 1 & 0 & -1 & 0 \\
		\hline
		$\eta$-ghost number & 0 & 0 & 0 & 0 & 0 & 0 & 0 & 0 & 0 & 0 & 0 & -1 & -1 & -1 & -1 & -1\\
		\hline
		$U(f)$ charge & 0 & 0 & 0 & 0 & -1 & -1 & 1 & 1 & 0 & 0 & 0 & 0 & 1 & 1 & -1 & -1 \\
		\hline
		Nature & F & B & F & B & B & F & F & B & B & B & F & F & B & F & B & F \\
		\hline
	\end{tabular}
	\caption{The quantum numbers of sources and their nature, i.e., bosonic (B) meaning that they are commuting and fermionic (F) that implies an anti-commuting source.}
	\label{Table 2}
\end{table}

\subsection{Slavnov-Taylor and other functional identities}

In this section, we list the set of symmetries enjoyed by the action $\Sigma$ \eqref{modtotact}. The first one to be listed is the functional identity which is the statement regarding the $Q$-invariance of $\Sigma$. This is translated into the so-called Slavnov-Taylor identity which, in the present case, could also be named extended Slavnov-Taylor identity:
\begin{itemize}
	\item The (extended) Slavnov-Taylor identity:
	\begin{equation}
		\mathcal{S}_Q(\Sigma) = 0\,,
	\end{equation}
	with
	\begin{equation}
		\begin{split}
			\mathcal{S}_Q(\Sigma) &= \int {\rm d}^3x\left(\frac{\delta \Sigma}{\delta \Omega^a_\mu}\frac{\delta \Sigma}{\delta A^a_\mu} +\frac{\delta \Sigma}{\delta L^a}\frac{\delta \Sigma}{\delta c^a} + \frac{\delta \Sigma}{\delta K^a}\frac{\delta \Sigma}{\delta \xi^a} + b^a\frac{\delta \Sigma}{\delta \bar{c}^a} + \omega^{ai}\frac{\delta \Sigma}{\delta \varphi^{ai}}\right.\\ 
			&+ \left. \bar{\varphi}^{ai}\frac{\delta \Sigma}{\delta \bar{\omega}^{ai}} + M^{ai}_\mu\frac{\delta \Sigma}{\delta N^{ai}_\mu} + U^{ai}_\mu\frac{\delta \Sigma}{\delta V^{ai}_\mu} + \sigma\frac{\delta \Sigma}{\delta \rho} + X^{i}\frac{\delta \Sigma}{\delta Y^{i}} - \bar{Y}^{abi}\frac{\delta \Sigma}{\delta \bar{X}^{abi}}\right) + \chi\frac{\partial \Sigma}{\partial \alpha}\,.
		\end{split}
	\end{equation}
	
There are other functional identities which are simply the classical equations of motions for some fields that can be extended to the quantum realm thanks to the quantum action principle (QAP) \cite{Piguet:1995er}. They are listed below:
	\item The gauge-fixing condition:
	\begin{equation}
		\frac{\delta \Sigma}{\delta b^a} = \partial_\mu A^a_\mu -  \alpha b^a - \frac{1}{2}\chi\bar{c}^a\,.
	\end{equation}
	\item The anti-ghost equation:
	\begin{equation}
		\frac{\delta \Sigma}{\delta \bar{c}^a} - \partial_\mu \frac{\delta \Sigma}{\delta \Omega^a_\mu} = \frac{1}{2}\chi b^a\,.
	\end{equation}
	\item The equations of motion of the Lagrange multiplier $\tau^a$:
	\begin{equation}
		\frac{\delta \Sigma}{\delta \tau^a} - \partial_\mu\frac{\delta \Sigma}{\delta \mathcal{J}^a_\mu} = 0\,.
	\end{equation}
	
As already mentioned, the multi-index $i=(a,\mu)$ introduces a very particular way of contracting a subset of fields and sources which carry a global charge. This entails a functional identity that encodes such an invariance:
	\item A global $U(f)$ symmetry:
	\begin{equation}
		U_{ij}\Sigma = 0\,,
	\end{equation}
	where
	\begin{equation}
		\begin{split}
			U_{ij} &= \int d^3x\left( \varphi^{ai}\frac{\delta }{\delta \varphi^{aj}} - \bar{\varphi}^{aj}\frac{\delta }{\delta \bar{\varphi}^{ai}} + \omega^{ai}\frac{\delta }{\delta \omega^{aj}} - \bar{\omega}^{aj}\frac{\delta }{\delta \bar{\omega}^{ai}} - M_\mu^{aj}\frac{\delta }{\delta M^{ai}_\mu} + V_\mu^{ai}\frac{\delta }{\delta V^{aj}_\mu}\right.\\
			&- \left. N_\mu^{aj}\frac{\delta }{\delta N^{ai}_\mu} + U_\mu^{ai}\frac{\delta }{\delta U^{aj}_\mu} + X^i\frac{\delta }{\delta X^j} + Y^i\frac{\delta }{\delta Y^j} - \bar{X}^{abj}\frac{\delta }{\delta \bar{X}^{abi}} - \bar{Y}^{abj}\frac{\delta }{\delta \bar{Y}^{abi}}\right)\,.
		\end{split}
	\end{equation}
	
There are other field equations that can be written in a compatible form with the QAP as follows:
	\item A set of linearly broken constraints:
	\begin{equation}
		\frac{\delta \Sigma}{\delta \bar{\varphi}^{ai}} + \partial_\mu\frac{\delta \Sigma}{\delta M^{ai}_\mu} + gf^{abc}V_\mu^{bi}\frac{\delta \Sigma}{\delta \mathcal{J}^c_\mu} = -\rho \varphi^{ai} + Y^i \eta^a\,,
	\end{equation}
	\begin{equation}
		\frac{\delta \Sigma}{\delta \varphi^{ai}} + \partial_\mu \frac{\delta \Sigma}{\delta V^{ai}_\mu} - gf^{abc}\bar{\varphi}^{bi}\frac{\delta \Sigma}{\delta \tau^c} + gf^{abc}M_\mu^{bi}\frac{\delta \Sigma}{\delta \mathcal{J}^c_\mu} = -\rho \bar{\varphi}^{ai} -\sigma\bar{\omega}^{ai} + \bar{Y}^{bai} \eta^b\,,
	\end{equation}
	\begin{equation}
		\frac{\delta \Sigma}{\delta \bar{\omega}^{ai}} + \partial_\mu\frac{\delta \Sigma}{\delta N^{ai}_\mu} - gf^{abc}U_\mu^{bi}\frac{\delta \Sigma}{\delta \mathcal{J}^c_\mu} = \rho \omega^{ai} + \sigma\varphi^{ai} - X^i \eta^a\,,
	\end{equation}
	\begin{equation}
		\frac{\delta \Sigma}{\delta \omega^{ai}} + \partial_\mu \frac{\delta \Sigma}{\delta U^{ai}_\mu} - gf^{abc}\bar{\omega}^{bi}\frac{\delta \Sigma}{\delta \tau^c} + gf^{abc}N_\mu^{bi}\frac{\delta \Sigma}{\delta \mathcal{J}^c_\mu} = -\rho \bar{\omega}^{ai} - \bar{X}^{bai} \eta^b\,.
	\end{equation}
	
The FP-ghost number as well as the $\eta$-ghost number conservation are encoded in two identities, consequences of two independent global symmetries, that are written as:
	\item FP-ghost number and $\eta$-ghost number Ward Identities:
	\begin{equation}
		\begin{split}
			&\int {\rm d}^3x \left( c^a\frac{\delta \Sigma}{\delta c^a} - \bar{c}^a\frac{\delta \Sigma}{\delta \bar{c}^a} + \omega^{ai}\frac{\delta \Sigma}{\delta \omega^{ai}} - \bar{\omega}^{ai}\frac{\delta \Sigma}{\delta \bar{\omega}^{ai}} - \Omega^a_\mu \frac{\delta \Sigma}{\delta \Omega^a_\mu} -2L^a\frac{\delta \Sigma}{\delta L^a} - K^a\frac{\delta \Sigma}{\delta K^a} \right.\\
			&+ \left. U^{ai}_\mu\frac{\delta \Sigma}{\delta U^{ai}_\mu} -  N^{ai}_\mu\frac{\delta \Sigma}{\delta N^{ai}_\mu} + X^i\frac{\delta \Sigma}{\delta X^i} - \bar{X}^{abi}\frac{\delta \Sigma}{\delta \bar{X}^{abi}} \right) + \chi \frac{\partial \Sigma}{\partial \chi} = 0\,,
		\end{split}
	\end{equation}
	\begin{equation}
		\int {\rm d}^3x \left( \eta^a\frac{\delta \Sigma}{\delta \eta^a} - \bar{\eta}^a\frac{\delta \Sigma}{\delta \bar{\eta}^a} -\Xi^a_\mu\frac{\delta \Sigma}{\delta \Xi^a_\mu} - X^i\frac{\delta \Sigma}{\delta X^i} - Y^i\frac{\delta \Sigma}{\delta Y^i} - \bar{X}^{abi}\frac{\delta \Sigma}{\delta \bar{X}^{abi}} - \bar{Y}^{abi}\frac{\delta \Sigma}{\delta \bar{Y}^{abi}} \right) = 0\,.
	\end{equation}
	
The standard localizing fields together with the sources that were introduced satisfy another symmetry that corresponds to an exchange of appropriate fields and sources with suitable signs. This will be named as in the standard literature as $\mathcal{R}_{ij}$ symmetry and the underlying functional identity reads:
	\item Exact $\mathcal{R}_{ij}$ symmetry:
	\begin{equation}
		\mathcal{R}_{ij}\Sigma = 0\,,
	\end{equation}
	where
	\begin{equation}
		\mathcal{R}_{ij} = \int {\rm d}^3x \left( \varphi^{ai}\frac{\delta}{\delta \omega^{aj}}  - \bar{\omega}^{aj}\frac{\delta}{\delta \bar{\varphi}^{ai}} + V^{ai}_\mu\frac{\delta}{\delta U^{aj}_\mu} - N^{aj}_\mu\frac{\delta}{\delta M^{ai}_\mu} + \bar{X}^{abj}\frac{\delta \Sigma}{\delta \bar{Y}^{abi}} + Y^i\frac{\delta \Sigma}{\delta X^j} \right)\,.
	\end{equation}
	
The auxiliary fields $(\bar{\eta},\eta)$ introduced in the localization procedure of the Stueckelberg-like field carry a very similar structure of the FP ghosts. In particular, since they are introduced to lift the determinant of the operator $\mathcal{M}^{ab}(A^h) = -\partial_\mu D^{ab}_{\mu} (A^h)$ from the path integral measure to the Botzmann weight and thanks to the transversality of $A^h_\mu$, we can write functional identities for $(\bar{\eta},\eta)$ very much similar to those that are present in the Landau gauge for the FP ghosts, i.e.:
	\item A local $\bar{\eta}^a$ equation:
	\begin{equation}
		\frac{\delta \Sigma}{\delta \bar{\eta}^a} - \partial_\mu \frac{\delta \Sigma}{\delta \Xi^a_\mu} = 0\,.
	\end{equation}
	\item An integrated, linearly broken $\eta^a$ equation:
	\begin{equation}
		\int {\rm d}^3x \left( \frac{\delta \Sigma}{\delta \eta^a} + gf^{abc}\bar{\eta}^b\frac{\delta \Sigma}{\delta \tau^c} -gf^{abc}\Xi^b_\mu\frac{\delta \Sigma}{\delta \mathcal{J}^c_\mu} \right) = \int {\rm d}^3x \left( +X^i \bar{\omega}^{ai} - Y^i\bar{\varphi}^{ai} + \bar{X}^{abi}\omega^{bi} - \bar{Y}^{abi}\varphi^{bi} \right)\,.
	\end{equation}
	
As it happens with the FP ghosts, we can write identities that relate the ghosts $(\bar{\eta},\eta)$ and the standard localizing ghosts in $(\bar{\omega},\omega)$:
	\item Localizing ghosts identities:
	\begin{equation}
		W_{(1)}^i \Sigma = \int {\rm d}^3x \left( \bar{\omega}^{ai}\frac{\delta \Sigma}{\delta \bar{\eta}^a} + \eta^a\frac{\delta \Sigma}{\delta \omega^{ai}} + N^{ai}_\mu \frac{\delta \Sigma}{\delta \Xi^a_\mu} + \rho\frac{\delta \Sigma}{\delta X^i} \right) = 0\,,
	\end{equation}
	\begin{equation}
		W_{(2)}^i \Sigma = \int {\rm d}^3x \left( \bar{\varphi}^{ai}\frac{\delta \Sigma}{\delta \bar{\eta}^a} - \eta^a\frac{\delta \Sigma}{\delta \varphi^{ai}} + M^{ai}_\mu\frac{\delta \Sigma}{\delta \Xi^a_\mu} - \rho\frac{\delta \Sigma}{\delta Y^i} + \sigma\frac{\delta \Sigma}{\delta X^i} \right) = 0\,,
	\end{equation}
	\begin{equation}
		W_{(3)}^i \Sigma = \int {\rm d}^3x \left( \varphi^{ai}\frac{\delta \Sigma}{\delta \bar{\eta}^a} - \eta^a\frac{\delta \Sigma}{\delta \bar{\varphi}^{ai}} - gf^{abc}\frac{\delta \Sigma}{\delta \bar{Y}^{abi}}\frac{\delta \Sigma}{\delta \tau^c} - V^{ai}_\mu\frac{\delta \Sigma}{\delta \Xi^a_\mu} + \rho\frac{\delta \Sigma}{\delta \bar{Y}^{aai}} \right) = 0\,,
	\end{equation}
	\begin{equation}
		W_{(4)}^i \Sigma = \int {\rm d}^3x \left( \omega^{ai}\frac{\delta \Sigma}{\delta \bar{\eta}^a} - \eta^a\frac{\delta \Sigma}{\delta \bar{\omega}^{ai}} + gf^{abc}\frac{\delta \Sigma}{\delta \bar{X}^{abi}}\frac{\delta \Sigma}{\delta \tau^c} + U^{ai}_\mu\frac{\delta \Sigma}{\delta \Xi^a_\mu} + \rho\frac{\delta \Sigma}{\delta \bar{X}^{aai}} + \sigma\frac{\delta \Sigma}{\delta \bar{Y}^{aai}} \right) = 0\,.
	\end{equation}
\end{itemize}
Those identities make explicit that the RGZ-YMCS action shares the same symmetries of the RGZ action. This is not surprising, since the addition of the Chern-Simons term is compatible with BRST invariance and does not involve any structure that arises from the elimination of (infinitesimal) Gribov copies. The identities above can be extended to the quantum action $\Gamma$, i.e., the generating functional of one-particle irreducible diagrams, thanks to the QAP, see \cite{Piguet:1995er}.

\section{Finiteness of the YMCS-RGZ theory \label{FYMCSRGZ}}

Before heading to the characterization of the invariant counterterm, we must take into account the power-counting renormalizability of YM theories (in the presence or not of the CS term) in 3 space-time dimensions. As stated before, they are not only renormalizable by power counting, but superrenormalizable, meaning that divergences appear only up to a finite order in the perturbative loop expansion. This can be seen from the superficial degree of divergence of one-particle irreducible Feynmann diagrams $\gamma$,
\begin{equation}\label{supdeg}
	d(\gamma) = 3 -\sum_{\Phi} d_\Phi N_\Phi -\frac{1}{2}N_g\,,
\end{equation}
with $N_\Phi$ being the number of amputated external legs of a given field $\Phi$ appearing in the diagram $\gamma$, $d_\Phi$ is the dimension of such field and $N_g$ is the power of the coupling constant $g$ in the loop integral associated with $\gamma$.
By treating the coupling constant $g$ as an external field of mass dimension 1/2 \cite{Maggiore:1995yd}, one can make use of the quantum action principle, so that (\ref{supdeg}) takes the same form as in a renormalizable theory, including $g$ in the summation over $\Phi$:
\begin{equation}
	d(\gamma) = 3 -\sum_{\Phi} d_\Phi N_\Phi\,.
\end{equation}
Generically, we can state that the invariant (i.e., compatible with all the symmetries of the underlying theory) counterterm is the most general integrated local quantity in the fields and sources, bounded by dimension 3 and of ghost numbers 0. But since the counterterm is generated by loop diagrams, they are always of order $g^2$ at least. Taking this into account, the dimension of the counterterm is bounded by 2. This means that the counterterm $\Gamma_{\rm CT}$ has the following structure,
\begin{equation}
	\Gamma_{\rm CT} = g^2\Sigma^{(1)}_{\rm CT} + g^4\Sigma^{(2)}_{\rm CT} + g^6\Sigma^{(3)}_{\rm CT} \,.
	\label{CT_Gen_form}
\end{equation}
The expression \eqref{CT_Gen_form} reveals the superrenormalizable nature of YMCS theories. If $g$ was dimensionless, then there would be no reason to truncate the expansion up to $g^6$. However since $g$ has mass dimension $1/2$, the dimensionality of the counterterm is satured at order $g^6$. At leading order, the invariant counterterm $\Sigma_{\rm CT}$ can be interpreted as a perturbation of the classical action $\Sigma$ (\ref{modtotact}), satisfying the properties aforementioned. Following the Algebraic Renormalization framework \cite{Piguet:1995er}, we demand that the perturbed action ($\Sigma+\epsilon\Sigma_{\rm CT}$) obeys the same symmetries as the classical action, up to order $O(\epsilon^2)$ terms in the expansion parameter $\epsilon$, which implies the following constraints:
\begin{equation}\label{ctcond}
	\begin{aligned}
		&\mathcal{B}_Q \Sigma_{\rm CT} = 0\,,\\
		&\frac{\delta \Sigma_{\rm CT}}{\delta b^a} = 0\,,\\
		&\frac{\delta \Sigma_{\rm CT}}{\delta \bar{c}^a} - \partial_\mu \frac{\delta \Sigma_{\rm CT}}{\delta \Omega^a_\mu} = 0\,,\\
		&\frac{\delta \Sigma_{\rm CT}}{\delta \tau^a} - \partial_\mu\frac{\delta \Sigma_{\rm CT}}{\delta \mathcal{J}^a_\mu} = 0\,,\\
		&U_{ij}\Sigma_{\rm CT} = 0\,,\\
		&\frac{\delta \Sigma_{\rm CT}}{\delta \bar{\varphi}^{ai}} + \partial_\mu\frac{\delta \Sigma_{\rm CT}}{\delta M^{ai}_\mu} + gf^{abc}V_\mu^{bi}\frac{\delta \Sigma_{\rm CT}}{\delta \mathcal{J}^c_\mu} = 0\,,\\
		&\frac{\delta \Sigma_{\rm CT}}{\delta \varphi^{ai}} + \partial_\mu \frac{\delta \Sigma_{\rm CT}}{\delta V^{ai}_\mu} - gf^{abc}\bar{\varphi}^{bi}\frac{\delta \Sigma_{\rm CT}}{\delta \tau^c} + gf^{abc}M_\mu^{bi}\frac{\delta \Sigma_{\rm CT}}{\delta \mathcal{J}^c_\mu} = 0\,,\\
		&\frac{\delta \Sigma_{\rm CT}}{\delta \bar{\omega}^{ai}} + \partial_\mu\frac{\delta \Sigma_{\rm CT}}{\delta N^{ai}_\mu} - gf^{abc}U_\mu^{bi}\frac{\delta \Sigma_{\rm CT}}{\delta \mathcal{J}^c_\mu} = 0\,,\\	
		&\frac{\delta \Sigma_{\rm CT}}{\delta \omega^{ai}} + \partial_\mu \frac{\delta \Sigma_{\rm CT}}{\delta U^{ai}_\mu} - gf^{abc}\bar{\omega}^{bi}\frac{\delta \Sigma_{\rm CT}}{\delta \tau^c} + gf^{abc}N_\mu^{bi}\frac{\delta \Sigma_{\rm CT}}{\delta \mathcal{J}^c_\mu} = 0\,,\\
		&\mathcal{R}_{ij}\Sigma_{\rm CT} = 0\,,\\		
		&\frac{\delta \Sigma_{\rm CT}}{\delta \bar{\eta}^a} - \partial_\mu \frac{\delta \Sigma_{\rm CT}}{\delta \Xi^a_\mu} = 0\,,\\
		&\int {\rm d}^3x \left( \frac{\delta \Sigma_{\rm CT}}{\delta \eta^a} + gf^{abc}\bar{\eta}^b\frac{\delta \Sigma_{\rm CT}}{\delta \tau^c} -gf^{abc}\Xi^b_\mu\frac{\delta \Sigma_{\rm CT}}{\delta \mathcal{J}^c_\mu} \right) = 0\,,\\
		&W_{(1,2,3,4)}^i \Sigma_{\rm CT} = 0\,.
	\end{aligned}
\end{equation}
The first condition of eqs.~(\ref{ctcond}) defines the linearized Slavnov-Taylor operator $\mathcal{B}_Q$, which is written as
	\begin{equation}\label{linearized}
	\begin{split}
		\mathcal{B}_Q &= \int {\rm d}^3x\left(\frac{\delta \Sigma}{\delta \Omega^a_\mu}\frac{\delta }{\delta A^a_\mu}+ \frac{\delta \Sigma}{\delta A^a_\mu}\frac{\delta }{\delta \Omega^a_\mu} +\frac{\delta \Sigma}{\delta L^a}\frac{\delta }{\delta c^a} + \frac{\delta \Sigma}{\delta c^a}\frac{\delta}{\delta L^a} + \frac{\delta \Sigma}{\delta K^a}\frac{\delta }{\delta \xi^a} + \frac{\delta \Sigma}{\delta \xi^a}\frac{\delta }{\delta K^a} + b^a\frac{\delta }{\delta \bar{c}^a} + \omega^{ai} \frac{\delta}{\delta \varphi^{ai}}\right.\\ 
		&+ \left. \bar{\varphi}^{ai}\frac{\delta }{\delta \bar{\omega}^{ai}} + M^{ai}_\mu\frac{\delta }{\delta N^{ai}_\mu} + U^{ai}_\mu\frac{\delta }{\delta V^{ai}_\mu} + \sigma\frac{\delta }{\delta \rho} + X^{i}\frac{\delta }{\delta Y^{i}} - \bar{Y}^{abi}\frac{\delta }{\delta \bar{X}^{abi}}\right) + \chi\frac{\partial }{\partial \alpha}\,,
	\end{split}
\end{equation}
with 
\begin{equation}
	\mathcal{B}_Q\mathcal{B}_Q = 0\,.
	\label{nilpotencyBQ}
\end{equation}
The general solution to such first condition of eqs.~(\ref{ctcond}), thanks to the nilpotency of the linearized Slavnov-Taylor operator, see \eqref{nilpotencyBQ}, is characterized by
\begin{equation}
	\Sigma_{\rm CT} = \Delta + \mathcal{B}_Q \Delta^{(-1)}\,.
\end{equation}
The functional $\Delta$ belongs to the cohomology of the linearized Slavnov-Taylor operator (\ref{linearized}), i.e., it is a nontrivial $Q$-cocycle.
The functional $\mathcal{B}_Q \Delta^{(-1)}$ is a trivial $Q$-cocyle.
Since the auxiliary fields and sources introduced are $Q$-doublets, they can only enter in the trivial cocyle. Moreover, $\Delta$ and $\Delta^{(-1)}$ should be integrated over three dimensions and their integrand must be local on fields and sources with ghost-number $0$ and $-1$, respectively. Upon the action of the linearized Slavnov-Taylor operator $\mathcal{B}_Q$, the resulting functional $\mathcal{B}_Q \Delta^{(-1)}$ has vanishing ghost number. 
The restriction of the cocyles to dimension 2, since they are at least of order $g^2$, see eq.~\eqref{CT_Gen_form}, leads to the following general expressions 
\begin{equation}
\Delta = \int {\rm d}^3x \Big[a_0 \epsilon_{\mu\rho\nu}\left(\frac{1}{2}A^a_\mu \partial_\rho A^a_\nu + \frac{g}{3!}f^{abc}A^a_\mu A^b_\rho A^c_\nu \right) + a_1 J + \lambda^{abcd} A^{h,a}_\mu A^{h,b}_\mu A^{h,c}_\nu A^{h,d}_\nu \Big]\,,
\label{NTCocyle}
\end{equation}
and
\begin{equation}
\Delta^{(-1)} = \int {\rm d}^3x~\epsilon_{\mu\rho\nu}\left[b_1 \bar{\omega}^{ab}_\mu \partial_{\rho} \varphi^{ab}_\nu + b_2 gf^{abd}\bar{\omega}^{ac}A^{h,d}_\rho\varphi^{bc}_\nu + b_3 N^{ab}_{\mu\rho} \varphi^{ab}_\nu + b_4 V^{ab}_{\mu\rho}\bar{\omega}^{ab}_\nu\right]\,.
\label{TCocycle}
\end{equation}
Next to that, the action of the linearized Slavnov-Taylor operator on \eqref{TCocycle} leads to
\begin{equation}
	\begin{split}
		\mathcal{B}_Q \Delta^{(-1)} = \int {\rm d}^3x~\epsilon_{\mu\rho\nu}&\left[b_1( \bar{\varphi}^{ac}_\mu \partial_{\rho} \varphi^{bc}_\nu - \bar{\omega}^{ac}_\mu \partial_{\rho} \omega^{bc}_\nu)+ b_2 gf^{abd}( \bar{\varphi}^{ac}_\mu A^{h,d}_\rho \varphi^{bc}_\nu - \bar{\omega}^{ac}_\mu A^{h,a}_\rho \omega^{bc}_\nu) \right.\\
		&+ \left. b_3 (M^{ab}_{\mu\rho} \omega^{ab}_\nu-N^{ab}_{\mu\rho} \omega^{ab}_\nu )+ b_4 (U^{ab}_{\mu\rho}\bar{\omega}^{ab}_\nu-V^{ab}_{\mu\rho}\bar{\varphi}^{ab}_\nu)\right]\,.
	\end{split}
\label{TCocycle2}
\end{equation}
It should be clear that the parameters $a_0$, $a_1$, $\lambda^{abcd}$, $b_1$, $b_2$, $b_3$ and $b_4$ all have mass dimension $1$. By enforcing the constraints \eqref{ctcond}, one gets a vanishing trivial cocycle, i.e., $b_1=b_2=b_3=b_4=0$. As for the non-trivial part, one should keep in mind that the coefficients $a_0$, $a_1$ and $\lambda^{abcd}$ do not depend on the gauge parameter $\alpha$ since it was introduced as a $Q$-doublet and hence only appears in the trivial cocycle. As such, the coefficients $a_0$, $a_1$ and $\lambda^{abcd}$ can be considered in a symplified setting where $\alpha = 0$, i.e., in the Landau gauge. In this situation, the dressed gauge field $A^{h,a}_\mu$ reduces to the standard gauge field $A^a_\mu$, see the Appendix of \cite{Capri:2017bfd}, thanks to the decoupling of the Stueckelberg-like field $\xi^a$. Therefore, the only admissible value of $\lambda^{abcd}$ is zero in order to preserve BRST (or rather, $Q$) invariance. Therefore, the general structure of the counterterm is 
\begin{equation}
	\Sigma_{\rm CT} = \int d^3x\left[\tilde{a}_0\, g^2\,\epsilon_{\mu\rho\nu}\left(\frac{1}{2}A_\mu^a \partial_\rho A_\nu^a + \frac{g}{3!}f^{abc} A_\mu^a A_\rho^a A_\nu^a\right)+\tilde{a}_1\, g^2 J\right]\,,
\end{equation}
where we have employed the following redefinitions, $a_0 = g^2 \tilde{a}_0$ and $a_1 = g^2 \tilde{a}_1$. At this stage, two comments are in order: The term proportional to the source $J$, when taken at the physical value of the source, is a simple additive term to the counterterm action. Hence, it does not affect the correlation functions of the theory and is harmless for the renormalization properties of the theory. As for the integrated term proportional to $\tilde{a}_0$, it is simply the Chern-Simons action. This term is not locally-invariant under BRST transformations, i.e., its invariance is only achieved in the integrated form. It is known that this kind of term does not contribute to the non-trivial part of the cohomology, see \cite{DelCima:1997pb,DelCima:1998bz,Barnich:1998ke}. Thus, we conclude that the counterterm action is trivial and, therefore, the YMCS theory, when quantized in linear covariant gauges in harmony with the elimination of infinitesimal Gribov copies, is finite. It is well-known that YMCS theories are finite and our results confirm that such a property remains when infinitesimal Gribov copies are eliminated and dynamical formation of condensates are taken into account. 

\section{Conclusions \label{Sect:Conc}} 

The quantization of Yang-Mills theories is a subtle issue. On the one hand, a purely gauge-invariant formulation is possible, but comes at the cost of using expensive numerical simulations and casts a shadow on the mechanisms that drives the outcomes that we measure by lattice simulations. Moreover, such a procedure involves a discretization of space(time) and recovering the continuum limit is a difficult task in general. On the other hand, the quantization in the continuum is typically tied to a gauge-fixing procedure. As discussed, non-Abelian gauge theories are plagued by the existence of Gribov ambiguities and a consistent removal or treatment of such spurious configurations is still an open problem. Nevertheless, it is fair to say that, over that past fourty years, we have understood several important properties associated with the elimination of infinitesimal Gribov copies in (linear) covariant gauges. The construction of the (refined) Gribov-Zwanziger setup in the Landau gauge is certainly a milestone in this endeavor. In particular, over the past ten years, an important issue within the RGZ framework was understood, i.e., the fate of BRST symmetry. In its original formulation in the Landau gauge, the RGZ action breaks BRST invariance explicitly but in a soft manner. Finding an appropriate prescription to restore BRST symmetry was crucial in order to construct an action that removes infinitesimal Gribov copies in linear covariant gauges in harmony with gauge-parameter independence. Yet the RGZ action is sufficiently complicated, and dealing with explicit computations and the associated renormalization in four dimensions is a challenging topic where first results are still being harvested \cite{deBrito:2023qfs,deBrito:2024ffa,Mintz:2017qri,Barrios:2024idr}.

Remarkably, the geometrical nature of the Gribov problem allows us to move to lower dimensions and deal with the same structure necessary to eliminate infinitesimal gauge copies. In three dimensions, however, the underlying theory is better behaved in the ultraviolet and, actually, gives finite results - a drastic simplification if compared to the four-dimensional case. Another remarkable property is that, in three dimensions, a mass of topological nature can be given to the gauge fields through the Chern-Simons term. Such a massive parameter competes with the dynamical mass generated by the elimination of Gribov copies, triggering a rich phase diagram that resembles a confining/deconfining transition. Up to date, the incorporation of the Chern-Simons term in the (R)GZ action was studied in the Landau gauge, in linear covariant gauges and in the maximal Abelian gauge, see \cite{Canfora:2013zza,Ferreira:2020kqv,Ferreira:2021gqh,Felix:2021eoq} and \cite{Gomez:2015aba} for the discussion in the presence of a Higgs-like field in the fundamental representation of the gauge group. Despite these investigations, one particular question remained unanswered: It is well-known that YMCS theories are finite \cite{DelCima:1998ur} in three-dimensions when quantized within the standard Faddeev-Popov procedure in the Landau gauge (recently, this has been extended to gauges that interpolate between covariant and non-covariant gauges, see \cite{Azevedo:2024cov}). Is this result affected by the elimination of infinitesimal Gribov copies? In the present work, we have verified that the removal of infinitesimal Gribov copies and the inclusion of condensates (of non-perturbative origin) do not spoil the finiteness of YMCS in linear covariant gauges. This was achieved by means of the Algebraic Renormalization framework, see \cite{Piguet:1995er}, which allows for a proof at all orders in perturbation theory and independent of the regularization scheme.

The present result paves the way for an investigation of the renormalization properties of YMCS-RGZ theories coupled to matter in linear covariant gauges, as well as a detailed study of the underlying renormalization properties of such theories quantized in the maximal Abelian gauge. Such investigations can play a very important role as toy models for the four-dimensional case. Due to the finiteness of the theory, this could open a great opportunity to understand qualitative properties of explicit computations of quantum corrections to correlation functions, without the intricacies of renormalization, in a theory free of infinitesimal Gribov copies. Thus, this can potentially help us in finding explicitly how those spurious configurations can affect physical quantities encoded in gauge-invariant correlation functions.

\section*{Acknowledgments} 

ADP acknowledges CNPq under the grant PQ-2 (312211/2022-8), FAPERJ under the ``Jovem Cientista do Nosso Estado'' program (E-26/205.924/2022). This work was partially supported by CAPES - Finance Code 001.


\bibliography{refs}

\end{document}